\documentclass[aps,10pt,final,
notitlepage, oneside, twocolumn, nobibnotes, nofootinbib,
superscriptaddress, showpacs, centertags, showkeys, assume]{revtex4}

\usepackage{graphicx}
\usepackage{amsmath}

\begin{document}

\title{Ground state $1/2^+$ octet baryon sum rules predicting
a chain of inequalities for hadron  photoproduction total
cross-sections on corresponding baryons}

\author{S.~Dubni\v{c}ka \footnote{e-mail: fyzidubn@savba.sk}} \affiliation{Inst.\ of Physics,
Slovak Academy of\ Sciences, D\'ubravsk\'a cesta 9, 845 11
Bratislava, Slovak Republic}
\author{A.Z.~Dubni\v ckov\'a\footnote{e-mail: dubnickova@fmph.uniba.sk}}\affiliation{Dept.\ of Theor.Physics, Comenius University,
842 48 Bratislava, Slovak Republic}
\author{E.~A.~Kuraev} \affiliation{Bogol'ubov\ Laboratory of Theoretical
Physics, JINR, Dubna, Russia}

\begin{abstract}
Sum rules  are derived relating Dirac mean square  radii and
anomalous magnetic moments of various couples of the ground state
$1/2^+$ octet  baryons   with the convergent integral of the
difference of hadron photoproduction cross-sections on the
corresponding baryons. Taking into account the present knowledge
of static parameters of baryons a chain of inequalities for total
hadronic photoproduction cross-sections on baryons is found from
those sum rules.
 \vspace{1pc}
\end{abstract}

 \maketitle

\section{INTRODUCTION}

Recently, the new sum rule has been derived \cite{Bartos04}
\begin{equation}\label{eq:1}
\frac{1}{3}\langle r_{1p}^2\rangle
-\frac{\kappa_{p}^2}{4m_{p}^2}+\frac{\kappa_{n}^2}{4m_{n}^2}=
\frac{2}{\pi^2\alpha}\int\limits_{\omega_{N}}^\infty
\frac{d\omega}{\omega} \big[\sigma_{tot}^{\gamma {p\to
X}}(\omega)- \sigma_{tot}^{\gamma {n\to X}}(\omega)\big]
\end{equation}
relating Dirac proton mean square radius $\langle r_{1p}^2\rangle$
and anomalous magnetic moments of proton $\kappa_p$ and neutron
$\kappa_n$ to the convergent integral over a difference of the
total proton and neutron photoproduction cross-sections, in which
a mutual cancelation of the rise of the corresponding
cross-sections for $\omega \to \infty$ ($\omega$ is the photon
energy in the laboratory frame), created by the Pomeron exchanges,
was achieved. Using similar ideas the new Cabibbo-Radicati
\cite{Cab66} like sum rules for various suitable couples of the
members of the pseudoscalar meson nonet have been found in Ref.
\cite{dubni06}. In this work, to be fascinated just by the very
precise satisfaction of the sum rule for a difference of proton
and neutron total photoproduction cross-sections evaluating both
sides of (\ref{eq:1}) (for more detail see \cite{Bartos04}) and
getting (1.93$\pm $ 0.18)mb and (1.92$\pm $ 0.32)mb, respectively,
we extend the method for a derivation of all possible sum rules
for various suitable couples of the members of the ground state
$1/2^+$ octet baryons.

The paper is organized as follows. The next section is devoted to a
brief derivation of Weizs{\"a}cker-Williams like relations between
differential baryon electroproduction and total hadron
photoproduction cross-sections. In Section III baryon sum rules in a
general form are derived on the basis of analytic properties of the
retarded part of a forward Compton scattering amplitude of virtual
photon on baryon. Applications to various couples of baryons are
carried out in Section IV. Conclusions are given in the last
Section.

\begin{figure*}[hbt]
\begin{center}
\includegraphics[scale=.7]{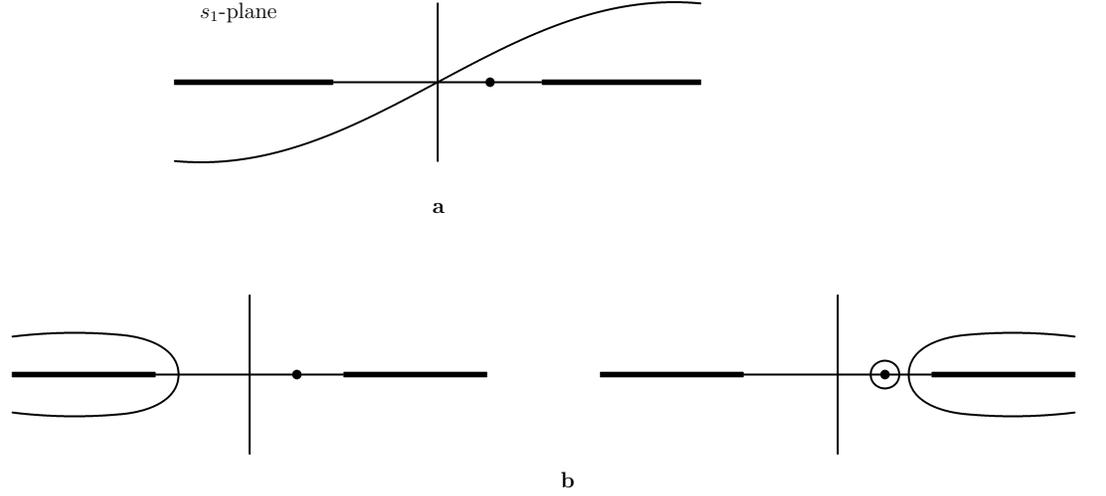}
\caption {\label{fig:1}Sum rule interpretation in $s_1$ plane.}
\end{center}
\end{figure*}

\section{WEIZS{\"A}CKER-WILLIAMS LIKE RELATIONS BETWEEN
DIFFERENTIAL BARYON ELECTROPRODUCTION AND TOTAL HADRON
PHOTOPRODUCTION CROSS-SECTIONS}

In a derivation of a such relation one considers currently a very
high energy peripheral electroproduction process on baryon $ B$
\begin{equation}
e^-(p_1) + B(p) \to e^-(p_1') + X, \label{a1}
\end{equation}
with the produced pure hadronic state $X$  moving closely to the
direction of the initial baryon, which is described by a matrix
element
\begin{equation}
M = i \frac{\sqrt{4\pi \alpha}}{q^2} \bar u(p_1^{'})\gamma_\mu
u(p_1) <X \mid J_\nu^{EM} \mid B>g^{\mu\nu},\label{a2}
\end{equation}
 in the one photon exchange approximation, where $m^2_X = (p+q)^2$ . Then the
Sudakov expansion \cite{Sud56} of the photon transferred four-vector
$q$
\begin{equation}\label{a3}
q=\beta_q \tilde{p}_1+\alpha_q\tilde{p}+q^{\bot}, \quad
q_{\bot}=(0,0,{\bf q}),\quad q_{\bot}^2=-{\bf q}^2
\end{equation}
into the almost light-like vectors
\begin{equation}\label{a4}
\tilde{p}_1=p_1-m_e^2p/(2p_1p), \quad
\tilde{p}=p-m_B^2p_1/(2p_1p)
\end{equation}
is applied and  the Gribov prescription  \cite{Grib70} for the
numerator of the photon Green function

\begin{equation}
g_{\mu\nu}=g_{\mu\nu}^{\bot}+\frac{2}{s}(\tilde{p}_{\mu}\tilde{p}_{1\nu}+\tilde{p}_{\nu}\tilde{p}_{1\mu})
 \approx\frac{2}{s} \tilde{p}_\mu\tilde{p}_{1\nu} \label{a5}
 \end{equation}
with $s=(p_1+p)^2\approx 2p_1p\gg Q^2 = -q^2$ is used  in (\ref{a2})
in order to write down for very high electron energy in (\ref{a1})
and small  photon momentum transfer squared $t=q^2=-Q^2=-{\bf q^2}$
the corresponding cross-section in the form
 \begin{eqnarray}\label{a6}
&&d\sigma^{e^-B\to e^-X}= \frac{4\pi\alpha}{s(q^2)^2}
p_1^{\mu}p_1^{\nu}\times \\ &&\sum_{X\neq B}\sum_{r=-1/2}^{1/2}
\langle B^{(r)}\mid J_\mu^{EM}\mid X\rangle^* \langle X \mid
J_\nu^{EM}\mid B^{(r)}\rangle d \Gamma \nonumber
\end{eqnarray}
with a summation through the created hadronic states $X$ and the
spin states of the initial baryon. If the relation $\int
d^{4}q\delta^{(4)}(p_1-q-p_1')=1$ is used in the phase space volume
$d\Gamma$ of the final particles one gets
\begin{equation}\label{a7}
d\Gamma = \frac{d s_1}{2s(2\pi)^3}d^2{\bf {q}}d\Gamma_X
\end{equation}
with
\begin{equation}
d\Gamma_X=(2\pi)^4
\delta^{(4)}(p+q-\sum_i^nq_i)\prod_i^n\frac{d^3q_i}{2E_i(2\pi)^3}\label{a8}
\end{equation}
\begin{equation}
 s_1=2(qp)=m_X^2+{\bf q^2}-m^2_{B}=s\beta_q.
 \label{a9}\end{equation}

Now the current conservation condition ($\alpha_q\tilde p$ gives a
negligible contribution)
\begin{eqnarray}\label{a10}
&&q^\mu \langle X\mid J_\mu^{EM}\mid B^{(r)}\rangle \approx \\
\nonumber&&\approx(\beta_q\tilde{p}_1+q_\bot)^\mu\langle X\mid
J_\mu^{EM}\mid B^{(r)}\rangle = 0,
\end{eqnarray}
is used in order to utilize in (\ref{a6}) the expression
\begin{eqnarray}\nonumber
\int p_1^\mu p_1^\nu \sum_{X\neq B}\sum_{r=-1/2}^{1/2}\langle
B^{(r)}\mid J_\mu^{EM}\mid X\rangle^* &&\\ \langle X \mid
J_\nu^{EM}\mid B^{(r)}\rangle d \Gamma_X=
 \label{a11} 2i\frac{s^2}{s_1^2}{\bf {q}^2} Im \tilde {A}^{(B)}(s_1,{\bf
 {q}}),&&
\end{eqnarray}
with the imaginary part of the retarded forward Compton scattering
amplitude $\tilde {A}^B(s_1, \bf q)$ \cite{Seco06} on a baryon. Then
for a difference of corresponding differential cross-sections of the
electroproduction on $B$ and $B'$ (after integration in (\ref{a6})
over $d\Gamma_X$, as well as over the invariant mass squared
$m_X^2$, i.e. over the variable $s_1$ to be interested only for
${\bf q}$ distribution) one finds
\begin{eqnarray} \nonumber
&&\Big(\frac{d\sigma^{e^-B\to e^-X}(s,{\bf q})}{d^2{\bf{q}}} -
\frac{d\sigma^{e^-B'\to
e^-X'}(s,{\bf q})}{d^2{\bf{q}}}\Big)=\\
\label{a12} &=&\frac{\alpha{\bf{q}^2}}
{4\pi^2}\int\limits_{s_1^{thr}}^\infty\frac{d
s_1}{s_1^2[{\bf{q}^2}+(m_es_1/s)^2]^2}\times \\
\nonumber&\times& [Im \tilde{A^B}(s_1,{\bf{q}})- Im
\tilde{A^{B'}}(s_1,{\bf{q}})].
\end{eqnarray}
Finally, if one neglects the second term in square brackets of the
denominator of the integral in (\ref{a12})(owing to the small value
of $m_e$ and high $s$ in comparison with $s_1$) and takes the limit
${\bf{q}^2}\to 0$ along with the expressions $d^2{\bf q}=\pi
d{\bf{q}}^2$ and $Im \tilde{A}^B(s_1,{\bf{q}})$=
$4s_1\sigma_{tot}^{\gamma^{*}B\to X}(s_1,{\bf{q}})$, one comes to
the Weizs{\"a}cker-Williams like relation
\begin{eqnarray} \label{a13}
&&{\bf{q}}^2\Big(\frac{d\sigma^{e^- B\to e^-X}}{d{\bf{q}}^2} -
\frac{d\sigma^{e^- B'\to e^-X'}}{d{\bf{q}^2}}\Big)_
{|_{{\bf{q}^2}\to 0}}=\\
\nonumber &=&\frac{\alpha}
{\pi}\int\limits_{s_1^{thr}}^\infty\frac{d s_1}{s_1}
[\sigma_{tot}^{\gamma B\to X}(s_1)-\sigma_{tot}^{\gamma B'\to
X}(s_1)]
\end{eqnarray}
relating the difference of ${\bf{q}^2}$-dependent differential
cross-sections of the processes (\ref{a1}) to the convergent
integral over the difference of the total hadron photoproduction
cross-sections on baryons.

\section{UNIVERSAL SUM RULE FOR OCTET BARYONS}

Universal sum rule is derived by exploiting the analytic
properties of the retarded Compton scattering amplitude
$\tilde{A}^B(s_1,{\bf{q}})$ in $s_1$- plane as presented in Fig.
1a, defining the integral $I$ over the path $C$ (for more detail
see \cite{Kura81}) in the $s_1$-plane
\begin{equation}\label{a14}
I=\int\limits_C d s_1 \frac{p_1^\mu p_1^\nu}{s^2}
\left(\tilde{A}^{B}_{\mu\nu}(s_1,{\bf{q}})-
\tilde{A}^{B'}_{\mu\nu}(s_1,{\bf{q}})\right )
\end{equation}
from the gauge invariant light-cone projection
$p_1^{\mu}p_1^{\nu}\tilde{A}_{\mu\nu}^B(s_1,{\bf{q}})$ of the
amplitude $\tilde{A}^B(s_1,{\bf{q}})$ and once closing the
contour $C$ to upper half-plane, another one to lower half-plane
(see Fig. 1b). As a result the following sum rule
\begin{eqnarray}\nonumber
& &\pi (Res^{B'}-Res^{B})
={\bf{q}}^2\int\limits_{r.h.}^\infty\frac{ds_1}
{s_1^2} [Im \tilde{A}^{B}(s_1,{\bf{q}})-\\
&-& Im \tilde{A}^{B'}(s_1,{\bf{q}})] \label{a15}
\end{eqnarray}
appears with (an averaging over the initial baryon and photon
spins is performed)
\begin{equation}\label{a16}
Res^{B}=2\pi\alpha( F^2_{1B}+ \frac{{\bf{q}^2}}{4m_B^2}F_{2B}^2)
\end{equation}
to be the residuum of the baryon intermediate state pole (see Fig.
1) contribution expressed through the Dirac and Pauli baryon
electromagnetic form factors and the left-hand cut contributions
expressed by an integral over the difference $[Im
\tilde{A}^{B}(s_1,{\bf{q}}) - Im \tilde{A}^{B'}(s_1,{\bf{q}})]$ are
assumed to be mutually annulated. Then, substituting (\ref{a16})
into (\ref{a15}) and  taking into account (\ref{a12}) from the
previous Section with $d^2{\bf q}=\pi d{\bf q^2}$, one comes to the
${\bf{q}^2}$ dependent baryon sum rule
\begin{eqnarray}
& &[F^2_{1B'}({-\bf{q^2}})-F^2_{1B'}(0)] -
[F^2_{1B}({\bf{-q^2}})-F^2_{1B}(0)] + \nonumber\\
&+&{\bf{q^2}}\big [\frac{F^2_{2B'}({\bf-{q^2}})}{4m_{B'}^2}-
\frac{F^2_{2B}({\bf-{q^2}})}{4m_{B}^2}\big ]=\nonumber \\
 &=&\frac{2}{\pi
\alpha^2}({\bf{q^2}})^2\Big(\frac{d\sigma^{e^- B\to
e^-X}}{d{\bf{q}}^2} - \frac{d\sigma^{e^- B'\to
e^-X}}{d{\bf{q}}^2}\Big ),\label{a17}
\end{eqnarray}
where the left-hand side was renormalized in order to separate the
pure strong interactions from electromagnetic ones. Besides,
substituting here for small values of ${\bf q^2}$ the relation
(\ref{a13}) from the previous Section,  using the laboratory
reference frame by $s_1=2m_B\omega$ and finally taking a derivative
according to ${\bf{q}^2}$ of both sides for ${\bf q^2}=0$, one gets
the universal octet baryon sum rule
\begin{eqnarray}\nonumber
&&\frac{1}{3}\big [F_{1B}(0)\langle r_{1B}^2
\rangle-F_{1B'}(0)\langle r_{1B'}^2 \rangle\big ]-
\big [\frac{\kappa_B^2}{4m_B^2}-\frac{\kappa_{B'}^2}{4m^2_{B'}}\big ]=\\
\label{a18}
&=&\frac{2}{\pi^2\alpha}\int\limits_{{\omega_B}}^{\infty}
\frac{d\omega} {\omega}\big[\sigma_{tot}^{\gamma B\to X}(\omega)-
\sigma_{tot}^{\gamma B'\to X}(\omega)\big]
\end{eqnarray}
relating Dirac baryon mean square radii $\langle r_{1B}^2\rangle$
and baryon anomalous magnetic moments $\kappa_B$ to the convergent
integral, in which a mutual cancelation of the rise of the
corresponding total cross-sections for $\omega\to\infty$ is
achieved.

\section{APPLICATION TO VARIOUS COUPLES OF OCTET BARYONS}

According to the SU(3) classification of existing hadrons there are
known the following members of the ground state $1/2^+$ baryon octet
($p$, $n$, $\Lambda^0$, $\Sigma^+$,  $\Sigma^0$, $\Sigma^-$,
$\Xi^0$, $\Xi^-$). As a result, by using the universal expression
(\ref{a18}) one can write down $8!/(2!(8-2)!)=28$ different sum
rules for total cross-sections of hadron photoproduction on ground
state $1/2^+$ octet baryons. The most precise of them (already
experimentally verified) proton-neutron sum rule (\ref{eq:1}) has
been previously published in \cite{Bartos04}. The rest of 27 sum
rules are explicitly presented in this paper.

If one considers couples of the isotriplet of $\Sigma$-hyperons and
separately couples of the isodoublet of $\Xi$-hyperons, one finds
\begin{eqnarray}\nonumber
&&\frac{1}{3}\big [\langle r_{1\Sigma^+}^2 \rangle -
\big [\frac{\kappa_{\Sigma^+}^2}{4m_{\Sigma^+}^2}-\frac{\kappa_{\Sigma^0}^2}{4m^2_{\Sigma^0}}\big ]=\\
\label{a19}
&=&\frac{2}{\pi^2\alpha}\int\limits_{\omega_{\Sigma^+}}^{\infty}
\frac{d\omega} {\omega}\big[\sigma_{tot}^{\gamma \Sigma^+\to
X}(\omega)- \sigma_{tot}^{\gamma \Sigma^0\to X}(\omega)\big ],
\end{eqnarray}
\begin{eqnarray}\nonumber
&&\frac{1}{3}\big [\langle r_{1\Sigma^+}^2 \rangle -\langle
r_{1\Sigma^-}^2 \rangle\big ]-
\big [\frac{\kappa_{\Sigma^+}^2}{4m_{\Sigma^+}^2}-\frac{\kappa_{\Sigma^-}^2}{4m^2_{\Sigma^-}}\big ]=\\
\label{a20}
&=&\frac{2}{\pi^2\alpha}\int\limits_{\omega_{\Sigma^+}}^{\infty}
\frac{d\omega} {\omega}\big[\sigma_{tot}^{\gamma \Sigma^+\to
X}(\omega)- \sigma_{tot}^{\gamma \Sigma^-\to X}(\omega)\big ],
\end{eqnarray}
\begin{eqnarray}\nonumber
&&\frac{1}{3}\langle r_{1\Sigma^-}^2 \rangle-
\big [\frac{\kappa_{\Sigma^0}^2}{4m_{\Sigma^0}^2}-\frac{\kappa_{\Sigma^-}^2}{4m^2_{\Sigma^-}}\big ]=\\
\label{a21}
&=&\frac{2}{\pi^2\alpha}\int\limits_{\omega_{\Sigma^0}}^{\infty}
\frac{d\omega} {\omega}\big[\sigma_{tot}^{\gamma \Sigma^0\to
X}(\omega)- \sigma_{tot}^{\gamma \Sigma^-\to X}(\omega)\big ],
\end{eqnarray}
and
\begin{eqnarray}\nonumber
&&\frac{1}{3}\langle r_{1\Xi^-}^2 \rangle-
\big [\frac{\kappa_{\Xi^0}^2}{4m_{\Xi^0}^2}-\frac{\kappa_{\Xi^-}^2}{4m^2_{\Xi^-}}\big ]=\\
\label{a22}
&=&\frac{2}{\pi^2\alpha}\int\limits_{\omega_{\Xi^0}}^{\infty}
\frac{d\omega} {\omega}\big[\sigma_{tot}^{\gamma \Xi^0\to
X}(\omega)- \sigma_{tot}^{\gamma \Xi^-\to X}(\omega)\big ],
\end{eqnarray}
respectively, which represent the second class of the baryon sum
rules what is concerned of their precision.

The third class of the 23 baryon sum rules is found by a
consideration of a couple of baryons always taken from different
isomultiplets of the ground state $1/2^+$ baryon octet and take
forms as follows
\begin{eqnarray}\nonumber
&&\frac{1}{3}\langle r_{1p}^2 \rangle-
\big [\frac{\kappa_{p}^2}{4m_{p}^2}-\frac{\kappa_{\Lambda^0}^2}{4m^2_{\Lambda^0}}\big ]=\\
\label{a23}
&=&\frac{2}{\pi^2\alpha}\int\limits_{\omega_{p}}^{\infty}
\frac{d\omega} {\omega}\big[\sigma_{tot}^{\gamma p\to X}(\omega)-
\sigma_{tot}^{\gamma \Lambda^0\to X}(\omega)\big ],
\end{eqnarray}

\begin{eqnarray}\nonumber
&&\frac{1}{3}\big [\langle r_{1p}^2 \rangle -\langle r_{1\Sigma^+}^2
\rangle\big ]-
\big [\frac{\kappa_{p}^2}{4m_{p}^2}-\frac{\kappa_{\Sigma^+}^2}{4m^2_{\Sigma^+}}\big ]=\\
\label{a24}
&=&\frac{2}{\pi^2\alpha}\int\limits_{\omega_{p}}^{\infty}
\frac{d\omega} {\omega}\big[\sigma_{tot}^{\gamma p \to X}(\omega)-
\sigma_{tot}^{\gamma \Sigma^+\to X}(\omega)\big ],
\end{eqnarray}

\begin{eqnarray}\nonumber
&&\frac{1}{3}\langle r_{1p}^2 \rangle-
\big [\frac{\kappa_{p}^2}{4m_{p}^2}-\frac{\kappa_{\Sigma^0}^2}{4m^2_{\Sigma^0}}\big ]=\\
\label{a25}
&=&\frac{2}{\pi^2\alpha}\int\limits_{\omega_{p}}^{\infty}
\frac{d\omega} {\omega}\big[\sigma_{tot}^{\gamma p\to X}(\omega)-
\sigma_{tot}^{\gamma \Sigma^0\to X}(\omega)\big ],
\end{eqnarray}

\begin{eqnarray}\nonumber
&&\frac{1}{3}\big [\langle r_{1p}^2 \rangle +\langle r_{1\Sigma^-}^2
\rangle\big ]-
\big [\frac{\kappa_{p}^2}{4m_{p}^2}-\frac{\kappa_{\Sigma^-}^2}{4m^2_{\Sigma^-}}\big ]=\\
\label{a26}
&=&\frac{2}{\pi^2\alpha}\int\limits_{\omega_{p}}^{\infty}
\frac{d\omega} {\omega}\big[\sigma_{tot}^{\gamma p \to X}(\omega)-
\sigma_{tot}^{\gamma \Sigma^-\to X}(\omega)\big ],
\end{eqnarray}

\begin{eqnarray}\nonumber
&&\frac{1}{3}\langle r_{1p}^2 \rangle-
\big [\frac{\kappa_{p}^2}{4m_{p}^2}-\frac{\kappa_{\Xi^0}^2}{4m^2_{\Xi^0}}\big ]=\\
\label{a27}
&=&\frac{2}{\pi^2\alpha}\int\limits_{\omega_{p}}^{\infty}
\frac{d\omega} {\omega}\big[\sigma_{tot}^{\gamma p\to X}(\omega)-
\sigma_{tot}^{\gamma \Xi^0\to X}(\omega)\big ],
\end{eqnarray}

\begin{eqnarray}\nonumber
&&\frac{1}{3}\big [\langle r_{1p}^2 \rangle +\langle r_{1\Xi^-}^2
\rangle\big ]-
\big [\frac{\kappa_{p}^2}{4m_{p}^2}-\frac{\kappa_{\Xi^-}^2}{4m^2_{\Xi^-}}\big ]=\\
\label{a28}
&=&\frac{2}{\pi^2\alpha}\int\limits_{\omega_{p}}^{\infty}
\frac{d\omega} {\omega}\big[\sigma_{tot}^{\gamma p \to X}(\omega)-
\sigma_{tot}^{\gamma \Xi^-\to X}(\omega)\big ],
\end{eqnarray}

\begin{eqnarray}\nonumber
&&-\big [\frac{\kappa_{n}^2}{4m_{n}^2}-\frac{\kappa_{\Lambda^0}^2}{4m^2_{\Lambda^0}}\big ]=\\
\label{a29}
&=&\frac{2}{\pi^2\alpha}\int\limits_{\omega_{n}}^{\infty}
\frac{d\omega} {\omega}\big[\sigma_{tot}^{\gamma n\to X}(\omega)-
\sigma_{tot}^{\gamma \Lambda^0\to X}(\omega)\big ],
\end{eqnarray}

\begin{eqnarray}\nonumber
&&-\frac{1}{3}\langle r_{1\Sigma^+}^2 \rangle-
\big [\frac{\kappa_{n}^2}{4m_{n}^2}-\frac{\kappa_{\Sigma^+}^2}{4m^2_{\Sigma^+}}\big ]=\\
\label{a30}
&=&\frac{2}{\pi^2\alpha}\int\limits_{\omega_{n}}^{\infty}
\frac{d\omega} {\omega}\big[\sigma_{tot}^{\gamma n\to X}(\omega)-
\sigma_{tot}^{\gamma \Sigma^+\to X}(\omega)\big ],
\end{eqnarray}

\begin{eqnarray}\nonumber
&&-\big [\frac{\kappa_{n}^2}{4m_{n}^2}-\frac{\kappa_{\Sigma^0}^2}{4m^2_{\Sigma^0}}\big ]=\\
\label{a31}
&=&\frac{2}{\pi^2\alpha}\int\limits_{\omega_{n}}^{\infty}
\frac{d\omega} {\omega}\big[\sigma_{tot}^{\gamma n\to X}(\omega)-
\sigma_{tot}^{\gamma \Sigma^0\to X}(\omega)\big ],
\end{eqnarray}

\begin{eqnarray}\nonumber
&&\frac{1}{3}\langle r_{1\Sigma^-}^2 \rangle-
\big [\frac{\kappa_{n}^2}{4m_{n}^2}-\frac{\kappa_{\Sigma^-}^2}{4m^2_{\Sigma^-}}\big ]=\\
\label{a32}
&=&\frac{2}{\pi^2\alpha}\int\limits_{\omega_{n}}^{\infty}
\frac{d\omega} {\omega}\big[\sigma_{tot}^{\gamma n\to X}(\omega)-
\sigma_{tot}^{\gamma \Sigma^-\to X}(\omega)\big ],
\end{eqnarray}

\begin{eqnarray}\nonumber
&&-\big [\frac{\kappa_{n}^2}{4m_{n}^2}-\frac{\kappa_{\Xi^0}^2}{4m^2_{\Xi^0}}\big ]=\\
\label{a33}
&=&\frac{2}{\pi^2\alpha}\int\limits_{\omega_{n}}^{\infty}
\frac{d\omega} {\omega}\big[\sigma_{tot}^{\gamma n\to X}(\omega)-
\sigma_{tot}^{\gamma \Xi^0\to X}(\omega)\big ],
\end{eqnarray}

\begin{eqnarray}\nonumber
&&\frac{1}{3}\langle r_{1\Xi^-}^2 \rangle-
\big [\frac{\kappa_{n}^2}{4m_{n}^2}-\frac{\kappa_{\Xi^-}^2}{4m^2_{\Xi^-}}\big ]=\\
\label{a34}
&=&\frac{2}{\pi^2\alpha}\int\limits_{\omega_{n}}^{\infty}
\frac{d\omega} {\omega}\big[\sigma_{tot}^{\gamma n\to X}(\omega)-
\sigma_{tot}^{\gamma \Xi^-\to X}(\omega)\big ],
\end{eqnarray}

\begin{eqnarray}\nonumber
&&-\frac{1}{3}\langle r_{1\Sigma^+}^2 \rangle-
\big [\frac{\kappa_{\Lambda^0}^2}{4m_{\Lambda^0}^2}-\frac{\kappa_{\Sigma^+}^2}{4m^2_{\Sigma^+}}\big ]=\\
\label{a35}
&=&\frac{2}{\pi^2\alpha}\int\limits_{\omega_{\Lambda^0}}^{\infty}
\frac{d\omega} {\omega}\big[\sigma_{tot}^{\gamma \Lambda^0\to
X}(\omega)- \sigma_{tot}^{\gamma \Sigma^+\to X}(\omega)\big ],
\end{eqnarray}

\begin{eqnarray}\nonumber
&&-\big [\frac{\kappa_{\Lambda^0}^2}{4m_{\Lambda^0}^2}-\frac{\kappa_{\Sigma^0}^2}{4m^2_{\Sigma^0}}\big ]=\\
\label{a36}
&=&\frac{2}{\pi^2\alpha}\int\limits_{\omega_{\Lambda^0}}^{\infty}
\frac{d\omega} {\omega}\big[\sigma_{tot}^{\gamma \Lambda^0\to
X}(\omega)- \sigma_{tot}^{\gamma \Sigma^0\to X}(\omega)\big ],
\end{eqnarray}

\begin{eqnarray}\nonumber
&&\frac{1}{3}\langle r_{1\Sigma^-}^2 \rangle-
\big [\frac{\kappa_{\Lambda^0}^2}{4m_{\Lambda^0}^2}-\frac{\kappa_{\Sigma^-}^2}{4m^2_{\Sigma^-}}\big ]=\\
\label{a37}
&=&\frac{2}{\pi^2\alpha}\int\limits_{\omega_{\Lambda^0}}^{\infty}
\frac{d\omega} {\omega}\big[\sigma_{tot}^{\gamma \Lambda^0\to
X}(\omega)- \sigma_{tot}^{\gamma \Sigma^-\to X}(\omega)\big ],
\end{eqnarray}

\begin{eqnarray}\nonumber
&&-\big [\frac{\kappa_{\Lambda^0}^2}{4m_{\Lambda^0}^2}-\frac{\kappa_{\Xi^0}^2}{4m^2_{\Xi^0}}\big ]=\\
\label{a38}
&=&\frac{2}{\pi^2\alpha}\int\limits_{\omega_{\Lambda^0}}^{\infty}
\frac{d\omega} {\omega}\big[\sigma_{tot}^{\gamma \Lambda^0\to
X}(\omega)- \sigma_{tot}^{\gamma \Xi^0\to X}(\omega)\big ],
\end{eqnarray}

\begin{eqnarray}\nonumber
&&\frac{1}{3}\langle r_{1\Xi^-}^2 \rangle-
\big [\frac{\kappa_{\Lambda^0}^2}{4m_{\Lambda^0}^2}-\frac{\kappa_{\Xi^-}^2}{4m^2_{\Xi^-}}\big ]=\\
\label{a39}
&=&\frac{2}{\pi^2\alpha}\int\limits_{\omega_{\Lambda^0}}^{\infty}
\frac{d\omega} {\omega}\big[\sigma_{tot}^{\gamma \Lambda^0\to
X}(\omega)- \sigma_{tot}^{\gamma \Xi^-\to X}(\omega)\big ],
\end{eqnarray}

\begin{eqnarray}\nonumber
&&\frac{1}{3}\langle r_{1\Sigma^+}^2 \rangle-
\big [\frac{\kappa_{\Sigma^+}^2}{4m_{\Sigma^+}^2}-\frac{\kappa_{\Xi^0}^2}{4m^2_{\Xi^0}}\big ]=\\
\label{a40}
&=&\frac{2}{\pi^2\alpha}\int\limits_{\omega_{\Sigma^+}}^{\infty}
\frac{d\omega} {\omega}\big[\sigma_{tot}^{\gamma \Sigma^+\to
X}(\omega)- \sigma_{tot}^{\gamma \Xi^0\to X}(\omega)\big ],
\end{eqnarray}

\begin{eqnarray}\nonumber
&&\frac{1}{3}\big [\langle r_{1\Sigma^+}^2 \rangle+\langle
r_{1\Xi^-}^2 \rangle\big ]-
\big [\frac{\kappa_{\Sigma^+}^2}{4m_{\Sigma^+}^2}-\frac{\kappa_{\Xi^-}^2}{4m^2_{\Xi^-}}\big ]=\\
\label{a41}
&=&\frac{2}{\pi^2\alpha}\int\limits_{\omega_{\Sigma^+}}^{\infty}
\frac{d\omega} {\omega}\big[\sigma_{tot}^{\gamma \Sigma^+\to
X}(\omega)- \sigma_{tot}^{\gamma \Xi^-\to X}(\omega)\big ],
\end{eqnarray}

\begin{eqnarray}\nonumber
&&-\big [\frac{\kappa_{\Sigma^0}^2}{4m_{\Sigma^0}^2}-\frac{\kappa_{\Xi^0}^2}{4m^2_{\Xi^0}}\big ]=\\
\label{a42}
&=&\frac{2}{\pi^2\alpha}\int\limits_{\omega_{\Sigma^0}}^{\infty}
\frac{d\omega} {\omega}\big[\sigma_{tot}^{\gamma \Sigma^0\to
X}(\omega)- \sigma_{tot}^{\gamma \Xi^0\to X}(\omega)\big ],
\end{eqnarray}

\begin{eqnarray}\nonumber
&&\frac{1}{3}\langle r_{1\Xi^-}^2 \rangle-
\big [\frac{\kappa_{\Sigma^0}^2}{4m_{\Sigma^0}^2}-\frac{\kappa_{\Xi^-}^2}{4m^2_{\Xi^-}}\big ]=\\
\label{a43}
&=&\frac{2}{\pi^2\alpha}\int\limits_{\omega_{\Sigma^0}}^{\infty}
\frac{d\omega} {\omega}\big[\sigma_{tot}^{\gamma \Sigma^0\to
X}(\omega)- \sigma_{tot}^{\gamma \Xi^-\to X}(\omega)\big ],
\end{eqnarray}

\begin{eqnarray}\nonumber
&&-\frac{1}{3}\langle r_{1\Sigma^-}^2 \rangle-
\big [\frac{\kappa_{\Sigma^-}^2}{4m_{\Sigma^-}^2}-\frac{\kappa_{\Xi^0}^2}{4m^2_{\Xi^0}}\big ]=\\
\label{a44}
&=&\frac{2}{\pi^2\alpha}\int\limits_{\omega_{\Sigma^-}}^{\infty}
\frac{d\omega} {\omega}\big[\sigma_{tot}^{\gamma \Sigma^-\to
X}(\omega)- \sigma_{tot}^{\gamma \Xi^0\to X}(\omega)\big ],
\end{eqnarray}

\begin{eqnarray}\nonumber
&&\frac{1}{3}\big [-\langle r_{1\Sigma^-}^2 \rangle+\langle
r_{1\Xi^-}^2 \rangle\big ]-
\big [\frac{\kappa_{\Sigma^-}^2}{4m_{\Sigma^-}^2}-\frac{\kappa_{\Xi^-}^2}{4m^2_{\Xi^-}}\big ]=\\
\label{a45}
&=&\frac{2}{\pi^2\alpha}\int\limits_{\omega_{\Sigma^-}}^{\infty}
\frac{d\omega} {\omega}\big[\sigma_{tot}^{\gamma \Sigma^-\to
X}(\omega)- \sigma_{tot}^{\gamma \Xi^-\to X}(\omega)\big ].
\end{eqnarray}

They are the most inaccurate in comparison with (\ref{eq:1}),
(\ref{a19})-(\ref{a22}) as there is a large difference in the masses
of joining pairs of particles and as a result there is no complete
annulation of the left hand cut contributions in (\ref{a15}).
\begin{table*}[t]
\caption{}
\begin{ruledtabular}
\begin{tabular}{c c c c c c c }\hline
B &  $I_B[mb]$ & $m_B [Gev]$ & $\kappa_B [\mu_N]$&$\langle r_{EB}^2
\rangle [fm^2]$&$3\kappa_B/2m_B^2 [fm^2]$&$\langle r_{1B}^2 \rangle
[fm^2]$\\  \hline
$p$  &0.9125         & 0.93827  & 1.7928 & 0.717& 0.119 & 0.598      \\
$n$ & 0.9100    & 0.93957    &-1.9130    &-0.113  &-0.127 & -0.240     \\
$\Lambda^0$  & 0.6454 &  1.11568   & -0.6130 & 0.110 & -0.029 & 0.081   \\
$\Sigma^+$ & 0.5679 & 1.18937 & 1.4580 & 0.600 & 0.060 & 0.660 \\
$\Sigma^0$& 0.5648 & 1.19264 & 0.6490    & -0.030  & 0.027 & -0.003     \\
$\Sigma^-$ & 0.5602   & 1.19745   & -0.1600   & 0.670 & -0.007 & 0.663   \\
$\Xi^0$  & 0.4647    & 1.31483 & -1.2500 & 0.130 & -0.042    & 0.088     \\
$\Xi^-$ &0.4601  & 1.32131 & 0.3493 & 0.490  & 0.012  & 0.502 \\
\end{tabular}
\end{ruledtabular}
\end{table*}

The latter follows from the careful analysis carried out in
\cite{Seco06}, where in analogy with QED the left-hand cut
contribution can be associated with contribution to the
cross-section of the process of baryon-antibaryon pair
electro-production on baryon $e^- B\to e^-2 B\bar B$, arising from
taking into account the identity of final state baryons. Keeping in
mind the minimal value of three baryon invariant mass squared
$s_{2min}= 8 m^2_B$, the left-hand cut contribution to the
derivative according to ${\bf q^2}$ for ${\bf q^2}=0$ entering the
derived sum rules have an order of magnitude
\begin{equation}
I_B=\frac{g^4 m_B^2}{(2\pi)^3 s_{2
min}^2}=\frac{g^4}{64(2\pi)^3m_B^2}, \quad \frac{g^2}{4\pi}=14.4
\label{a46}
\end{equation}
which depends on the baryon mass squared and so, the error in
derived sum rules increases with the increased difference in the
masses of joining pairs of baryons (see Table I). Now it is clear
that the proton-neutron sum rule is the most precise and then follow
the sum rules $\Sigma^+ -\Sigma^0$, $\Sigma^+ -\Sigma^-$,
$\Sigma^0-\Sigma^-$ and $\Xi^0-\Xi^-$.

In order to evaluate the left hand sides of the derived sum rules
(\ref{a19})-(\ref{a45}) and to draw out some phenomenological
consequences, one needs the reliable values of Dirac baryon mean
square radii $\langle r_{1B}^2 \rangle$ and baryon anomalous
magnetic moments $\kappa_B$. The latter are known (besides
$\Sigma^0$, which is found from the well known relation
$\kappa_{\Sigma^+}+ \kappa_{\Sigma^-}$=$2\kappa_{\Sigma^0}$)
experimentally (see the third column in Table I),
however, to calculate $\langle r_{1B}^2 \rangle$ by
means of the difference of the baryon electric mean square radius
$\langle r_{EB}^2 \rangle$ and Foldy term (well known for all ground
state octet baryons from the experimental information on the
magnetic moments \cite{Rev06})
\begin{equation}
\langle r_{1B}^2 \rangle =\langle r_{EB}^2
\rangle-\frac{3\kappa_B}{2 m_B^2}, \label{a47}
\end{equation}
we are in need of the reliable values of $\langle r_{EB}^2
\rangle$. They are known experimentally only for the proton,
neutron and $\Sigma^-$-hyperon \cite{Rev06}. Fortunately there are
recent results of Kubis and Meissner \cite{Meiss01} to fourth
order in relativistic baryon chiral perturbation theory (giving
predictions for the $\Sigma^-$ charge radius and the
$\Lambda$-$\Sigma^0$ transition moment in excellent agreement with
the available experimental information), which solve our problem
completely. All necessary information from \cite{Rev06} and
\cite{Meiss01} is collected in Table I, where also numerical
values of corresponding $\langle r^2_{1B} \rangle$ are presented.
%\begin{widetext}

 Calculating the left-hand side of all sum rules one finds
{ \begin{widetext}

\begin{equation}
\label{a48} \frac{2}{\pi^2\alpha}\int\limits_{\omega_{p}}^{\infty}
\frac{d\omega} {\omega}\big[\sigma_{tot}^{\gamma p\to X}(\omega)-
\sigma_{tot}^{\gamma n\to X}(\omega)\big ]=2.0415
  {\textrm mb},\quad \textrm{
  then in averaged}\quad
\sigma_{tot}^{\gamma p \to X}(\omega)> \sigma_{tot}^{\gamma n\to
X}(\omega)
\end{equation}

\begin{equation}
\label{a49}
\frac{2}{\pi^2\alpha}\int\limits_{\omega_{\Sigma^+}}^{\infty}
\frac{d\omega} {\omega}\big[\sigma_{tot}^{\gamma \Sigma^+\to
X}(\omega)- \sigma_{tot}^{\gamma \Sigma^0\to X}(\omega)\big ]=2.0825
  {\textrm mb},\quad \textrm{
  then in averaged}\quad
\sigma_{tot}^{\gamma \Sigma^+\to X}(\omega)> \sigma_{tot}^{\gamma
\Sigma^0\to X}(\omega)
\end{equation}

\begin{equation}
\label{a50}
\frac{2}{\pi^2\alpha}\int\limits_{\omega_{\Sigma^+}}^{\infty}
\frac{d\omega} {\omega}\big[\sigma_{tot}^{\gamma \Sigma^+\to
X}(\omega)- \sigma_{tot}^{\gamma \Sigma^-\to X}(\omega) \big
]=4.2654  {\textrm mb}, \quad \textrm{
  then in averaged}\quad \sigma_{tot}^{\gamma \Sigma^+\to
X}(\omega)> \sigma_{tot}^{\gamma \Sigma^-\to X}(\omega)
\end{equation}

\begin{equation}
\label{a51}
\frac{2}{\pi^2\alpha}\int\limits_{\omega_{\Sigma^0}}^{\infty}
\frac{d\omega} {\omega}\big[\sigma_{tot}^{\gamma \Sigma^0\to
X}(\omega)- \sigma_{tot}^{\gamma \Sigma^-\to X}(\omega)\big ]=
2.1829  {\textrm mb}, \quad \textrm{
  then in averaged}\quad \sigma_{tot}^{\gamma \Sigma^0\to
X}(\omega)> \sigma_{tot}^{\gamma \Sigma^-\to X}(\omega)
\end{equation}

\begin{equation}
\label{a52}
\frac{2}{\pi^2\alpha}\int\limits_{\omega_{\Xi^0}}^{\infty}
\frac{d\omega} {\omega}\big[\sigma_{tot}^{\gamma \Xi^0\to
X}(\omega)- \sigma_{tot}^{\gamma \Xi^-\to X}(\omega)\big ]=1.5921
    {\textrm mb}, \quad \textrm{then in averaged}\quad \sigma_{tot}^{\gamma
\Xi^0\to X}(\omega)> \sigma_{tot}^{\gamma \Xi^-\to X}(\omega)
\end{equation}

\begin{equation}
\label{a53} \frac{2}{\pi^2\alpha}\int\limits_{\omega_{p}}^{\infty}
\frac{d\omega} {\omega}\big[\sigma_{tot}^{\gamma p\to X}(\omega)-
\sigma_{tot}^{\gamma \Lambda^0\to X}(\omega)\big ]=1.6673   {\textrm
mb},\quad \textrm{then in averaged}\quad \sigma_{tot}^{\gamma p\to
X}(\omega)> \sigma_{tot}^{\gamma \Lambda^0\to X}(\omega)
\end{equation}

\begin{equation}
\label{a54} \frac{2}{\pi^2\alpha}\int\limits_{\omega_{p}}^{\infty}
\frac{d\omega} {\omega}\big[\sigma_{tot}^{\gamma p \to X}(\omega)-
\sigma_{tot}^{\gamma \Sigma^+\to X}(\omega)\big ]=-0.4158   {\textrm
mb},\quad \textrm{then in averaged}\quad \sigma_{tot}^{\gamma p \to
X}(\omega)<\sigma_{tot}^{\gamma \Sigma^+\to X}(\omega)
\end{equation}

\begin{equation}
\label{a55} \frac{2}{\pi^2\alpha}\int\limits_{\omega_{p}}^{\infty}
\frac{d\omega} {\omega}\big[\sigma_{tot}^{\gamma p\to X}(\omega)-
\sigma_{tot}^{\gamma \Sigma^0\to X}(\omega)\big ]=1.6667  {\textrm
mb},\quad \textrm{then in averaged}\quad \sigma_{tot}^{\gamma p\to
X}(\omega)> \sigma_{tot}^{\gamma \Sigma^0\to X}(\omega)
\end{equation}

\begin{equation}
\label{a56} \frac{2}{\pi^2\alpha}\int\limits_{\omega_{p}}^{\infty}
\frac{d\omega} {\omega}\big[\sigma_{tot}^{\gamma p \to X}(\omega)-
\sigma_{tot}^{\gamma \Sigma^-\to X}(\omega)\big ]= 3.8496  {\textrm
mb},\quad \textrm{then in averaged}\quad \sigma_{tot}^{\gamma p \to
X}(\omega)> \sigma_{tot}^{\gamma \Sigma^-\to X}(\omega)
\end{equation}

\begin{equation}
\label{a57} \frac{2}{\pi^2\alpha}\int\limits_{\omega_{p}}^{\infty}
\frac{d\omega} {\omega}\big[\sigma_{tot}^{\gamma p\to X}(\omega)-
\sigma_{tot}^{\gamma \Xi^0\to X}(\omega)\big ]=1.7259  {\textrm
mb},\quad \textrm{then in averaged}\quad \sigma_{tot}^{\gamma p\to
X}(\omega)> \sigma_{tot}^{\gamma \Xi^0\to X}(\omega)
\end{equation}

\begin{equation}
\label{a58} \frac{2}{\pi^2\alpha}\int\limits_{\omega_{p}}^{\infty}
\frac{d\omega} {\omega}\big[\sigma_{tot}^{\gamma p \to X}(\omega)-
\sigma_{tot}^{\gamma \Xi^-\to X}(\omega)\big ]=3.3180   {\textrm
mb},\quad \textrm{then in averaged}\quad \sigma_{tot}^{\gamma p \to
X}(\omega)> \sigma_{tot}^{\gamma \Xi^-\to X}(\omega)
\end{equation}

\begin{equation}
\label{a59} \frac{2}{\pi^2\alpha}\int\limits_{\omega_{n}}^{\infty}
\frac{d\omega} {\omega}\big[\sigma_{tot}^{\gamma n\to X}(\omega)-
\sigma_{tot}^{\gamma \Lambda^0\to X}(\omega)\big ]=-0.3260
{\textrm mb},\quad \textrm{then in averaged}\quad
\sigma_{tot}^{\gamma n\to X}(\omega)< \sigma_{tot}^{\gamma
\Lambda^0\to X}(\omega)
\end{equation}

\begin{equation}
\label{a60} \frac{2}{\pi^2\alpha}\int\limits_{\omega_{n}}^{\infty}
\frac{d\omega} {\omega}\big[\sigma_{tot}^{\gamma n\to X}(\omega)-
\sigma_{tot}^{\gamma \Sigma^+\to X}(\omega)\big ]=-2.4573   {\textrm
mb},\quad \textrm{then in averaged}\quad \sigma_{tot}^{\gamma n\to
X}(\omega)< \sigma_{tot}^{\gamma \Sigma^+\to X}(\omega)
\end{equation}

\begin{equation}
\label{a61} \frac{2}{\pi^2\alpha}\int\limits_{\omega_{n}}^{\infty}
\frac{d\omega} {\omega}\big[\sigma_{tot}^{\gamma n\to X}(\omega)-
\sigma_{tot}^{\gamma \Sigma^0\to X}(\omega)\big ]=-0.3747 {\textrm
mb},\quad \textrm{then in averaged}\quad \sigma_{tot}^{\gamma n\to
X}(\omega)< \sigma_{tot}^{\gamma \Sigma^0\to X}(\omega)
\end{equation}

\begin{equation}
\label{a62} \frac{2}{\pi^2\alpha}\int\limits_{\omega_{n}}^{\infty}
\frac{d\omega} {\omega}\big[\sigma_{tot}^{\gamma n\to X}(\omega)-
\sigma_{tot}^{\gamma \Sigma^-\to X}(\omega)\big ]= 1.8082  {\textrm
mb},\quad \textrm{then in averaged}\quad \sigma_{tot}^{\gamma n\to
X}(\omega)> \sigma_{tot}^{\gamma \Sigma^-\to X}(\omega)
\end{equation}

\begin{equation}
\label{a63} \frac{2}{\pi^2\alpha}\int\limits_{\omega_{n}}^{\infty}
\frac{d\omega} {\omega}\big[\sigma_{tot}^{\gamma n\to X}(\omega)-
\sigma_{tot}^{\gamma \Xi^0\to X}(\omega)\big ]= -0.3156  {\textrm
mb},\quad \textrm{then in averaged}\quad \sigma_{tot}^{\gamma n\to
X}(\omega)< \sigma_{tot}^{\gamma \Xi^0\to X}(\omega)
\end{equation}

\begin{equation}
\label{a64} \frac{2}{\pi^2\alpha}\int\limits_{\omega_{n}}^{\infty}
\frac{d\omega} {\omega}\big[\sigma_{tot}^{\gamma n\to X}(\omega)-
\sigma_{tot}^{\gamma \Xi^-\to X}(\omega)\big ]=1.2766  {\textrm
mb},\quad \textrm{then in averaged}\quad \sigma_{tot}^{\gamma n\to
X}(\omega)> \sigma_{tot}^{\gamma \Xi^-\to X}(\omega)
\end{equation}

\begin{equation}
\label{a65}
\frac{2}{\pi^2\alpha}\int\limits_{\omega_{\Lambda^0}}^{\infty}
\frac{d\omega} {\omega}\big[\sigma_{tot}^{\gamma \Lambda^0\to
X}(\omega)- \sigma_{tot}^{\gamma \Sigma^+\to X}(\omega)\big ]=
-2.0831 {\textrm mb},\quad \textrm{then in averaged}\quad
\sigma_{tot}^{\gamma \Lambda^0\to X}(\omega)< \sigma_{tot}^{\gamma
\Sigma^+\to X}(\omega)
\end{equation}

\begin{equation}
\label{a66}
\frac{2}{\pi^2\alpha}\int\limits_{\omega_{\Lambda^0}}^{\infty}
\frac{d\omega} {\omega}\big[\sigma_{tot}^{\gamma \Lambda^0\to
X}(\omega)- \sigma_{tot}^{\gamma \Sigma^0\to X}(\omega)\big ]=
-0.0006   {\textrm mb},\quad \textrm{then in averaged}\quad
\sigma_{tot}^{\gamma \Lambda^0\to X}(\omega)\approx
\sigma_{tot}^{\gamma \Sigma^0\to X}(\omega)
\end{equation}

\begin{equation}
\label{a67}
\frac{2}{\pi^2\alpha}\int\limits_{\omega_{\Lambda^0}}^{\infty}
\frac{d\omega} {\omega}\big[\sigma_{tot}^{\gamma \Lambda^0\to
X}(\omega)- \sigma_{tot}^{\gamma \Sigma^-\to X}(\omega)\big ]=2.1823
  {\textrm{mb}},\quad \textrm{then in averaged}\quad \sigma_{tot}^{\gamma
\Lambda^0\to X}(\omega)> \sigma_{tot}^{\gamma \Sigma^-\to X}(\omega)
\end{equation}

\begin{equation}
\label{a68}
\frac{2}{\pi^2\alpha}\int\limits_{\omega_{\Lambda^0}}^{\infty}
\frac{d\omega} {\omega}\big[\sigma_{tot}^{\gamma \Lambda^0\to
X}(\omega)- \sigma_{tot}^{\gamma \Xi^0\to X}(\omega)\big ]=0.0586
{\textrm{mb}},\quad \textrm{then in averaged}\quad
\sigma_{tot}^{\gamma \Lambda^0\to X}(\omega)> \sigma_{tot}^{\gamma
\Xi^0\to X}(\omega)
\end{equation}

\begin{equation}
\label{a69}
\frac{2}{\pi^2\alpha}\int\limits_{\omega_{\Lambda^0}}^{\infty}
\frac{d\omega} {\omega}\big[\sigma_{tot}^{\gamma \Lambda^0\to
X}(\omega)- \sigma_{tot}^{\gamma \Xi^-\to X}(\omega)\big ]=2.1823
{\textrm{mb}} ,\quad \textrm{then in averaged}\quad
\sigma_{tot}^{\gamma \Lambda^0\to X}(\omega)> \sigma_{tot}^{\gamma
\Xi^-\to X}(\omega)
\end{equation}

\begin{equation}
\label{a70}
\frac{2}{\pi^2\alpha}\int\limits_{\omega_{\Sigma^+}}^{\infty}
\frac{d\omega} {\omega}\big[\sigma_{tot}^{\gamma \Sigma^+\to
X}(\omega)- \sigma_{tot}^{\gamma \Xi^0\to X}(\omega)\big ]=2.1417
\textrm{mb},\quad \textrm{then in averaged}\quad
\sigma_{tot}^{\gamma \Sigma^+\to X}(\omega)> \sigma_{tot}^{\gamma
\Xi^0\to X}(\omega)
\end{equation}

\begin{equation}
\label{a71}
\frac{2}{\pi^2\alpha}\int\limits_{\omega_{\Sigma^+}}^{\infty}
\frac{d\omega} {\omega}\big[\sigma_{tot}^{\gamma \Sigma^+\to
X}(\omega)- \sigma_{tot}^{\gamma \Xi^-\to X}(\omega)\big ]=3.7338
\textrm{mb},\quad \textrm{then in averaged}\quad
\sigma_{tot}^{\gamma \Sigma^+\to X}(\omega)> \sigma_{tot}^{\gamma
\Xi^-\to X}(\omega)
\end{equation}

\begin{equation}
\label{a72}
\frac{2}{\pi^2\alpha}\int\limits_{\omega_{\Sigma^0}}^{\infty}
\frac{d\omega} {\omega}\big[\sigma_{tot}^{\gamma \Sigma^0\to
X}(\omega)- \sigma_{tot}^{\gamma \Xi^0\to X}(\omega)\big ]=0.1168
\textrm{mb},\quad \textrm{then in averaged}\quad
\sigma_{tot}^{\gamma \Sigma^0\to X}(\omega)> \sigma_{tot}^{\gamma
\Xi^0\to X}(\omega)
\end{equation}

\begin{equation}
\label{a73}
\frac{2}{\pi^2\alpha}\int\limits_{\omega_{\Sigma^0}}^{\infty}
\frac{d\omega} {\omega}\big[\sigma_{tot}^{\gamma \Sigma^0\to
X}(\omega)- \sigma_{tot}^{\gamma \Xi^-\to X}(\omega)\big ]=1.5732
\textrm{mb},\quad \textrm{then in averaged}\quad
\sigma_{tot}^{\gamma \Sigma^0\to X}(\omega)> \sigma_{tot}^{\gamma
\Xi^\to X}(\omega)
\end{equation}

\begin{equation}
\label{a74}
\frac{2}{\pi^2\alpha}\int\limits_{\omega_{\Sigma^-}}^{\infty}
\frac{d\omega} {\omega}\big[\sigma_{tot}^{\gamma \Sigma^-\to
X}(\omega)- \sigma_{tot}^{\gamma \Xi^0\to X}(\omega)\big ]=-2.1238
\textrm{mb},\quad \textrm{then in averaged}\quad
\sigma_{tot}^{\gamma \Sigma^-\to X}(\omega)< \sigma_{tot}^{\gamma
\Xi^0\to X}(\omega)
\end{equation}

\begin{equation}
\label{a75}
\frac{2}{\pi^2\alpha}\int\limits_{\omega_{\Sigma^-}}^{\infty}
\frac{d\omega} {\omega}\big[\sigma_{tot}^{\gamma \Sigma^-\to
X}(\omega)- \sigma_{tot}^{\gamma \Xi^-\to X}(\omega)\big ]=-0.5316
\textrm{mb},\quad \textrm{then in averaged}\quad
\sigma_{tot}^{\gamma \Sigma^-\to X}(\omega)< \sigma_{tot}^{\gamma
\Xi^-\to X}(\omega),
\end{equation}
\end{widetext}}

from where one gets the following chain of inequalities
{\begin{widetext}
\begin{equation}
\sigma_{tot}^{\gamma \Sigma^+\to X}(\omega)>
\sigma_{tot}^{\gamma p\to X}(\omega)> \sigma_{tot}^{\gamma
\Lambda^0\to X}(\omega)
\approx   \sigma_{tot}^{\gamma \Sigma^0\to X}(\omega)>
\sigma_{tot}^{\gamma \Xi^0\to X}(\omega)>
 \sigma_{tot}^{\gamma n\to X}(\omega)>
\sigma_{tot}^{\gamma \Xi^-\to X}(\omega)>\sigma_{tot}^{\gamma
\Sigma^-\to X}(\omega)
\end{equation}
\end{widetext}}
for total cross-sections of hadron photoproduction on ground state
$1/2^+$ octet baryons.

Experimental tests of the derived sum rules could be practically
carried out provided there exist total hadron photoproduction
cross-sections on hyperons as a function of energy, which are,
however, missing till now. Nevertheless, the idea of intensive
photon beams generated by the electron beams of linear $e^+e^-$
colliders by using the process of the backward Compton scattering of
laser light off the energy electrons \cite{Ginzb81} is encouraging
and one expects in near future that the measurements of the total
hadron photoproduction cross-sections on hyperons can be practically
achievable.

\section{CONCLUSIONS}

Considering the very high energy peripheral electron-baryon
scattering with a production of a hadronic state $X$ moving closely
to the direction of initial baryon, then exploiting analytic
properties of the forward Compton scattering amplitude on the same
baryon, for the case of small transferred momenta new sum rules,
relating  Dirac baryon mean square radii and the baryon anomalous
magnetic moments with the convergent integrals over a difference of
total hadron photoproduction cross-sections on baryons are derived.
Evaluating the left-hand sides of the sum rules the chain of
inequalities for total cross-sections of hadron photoproduction on
all ground state $1/2^+$ octet baryons has been found. Possible
practical tests of the derived sum rules are discussed as well.

\bigskip

The work was partly supported by Slovak Grant Agency for Sciences
VEGA, Grant No. 2/4099/26 (S.D. and A.Z.D).  A.Z. Dubni\v ckova
would like to thank University di Trieste for warm hospitality at
the early stage of this work, and Professor N.Paver for numerous
discussions.

\end{document}